\documentclass[12pt]{article}
\pdfoutput=1
\usepackage{amsmath}
\usepackage{amssymb}
\usepackage{graphicx}
\usepackage{subcaption}
\usepackage{url}
\usepackage{slashed}
\usepackage{arydshln}
\usepackage[sort&compress, numbers, merge]{natbib}
\usepackage{multirow}

\usepackage[utf8]{inputenc}

\def\bea{\begin{eqnarray}}
\def\eea{\end{eqnarray}}

\usepackage[colorlinks=true, linkcolor=blue, citecolor=blue,
urlcolor=black]{hyperref}

\begin{document}

\newcommand{\TeV}{\text{TeV}}
\newcommand{\GeV}{\text{GeV}}
\newcommand{\MeV}{\text{MeV}}
\newcommand{\keV}{\text{keV}}
\newcommand{\eV}{\text{eV}}

\newcommand{\MET}{E_{\rm T}^{\text{miss}}}

\newcommand{\lrf}[2]{ \left(\frac{#1}{#2}\right)}
\newcommand{\lrfp}[3]{ \left(\frac{#1}{#2} \right)^{#3}}
\newcommand{\vev}[1]{\left\langle #1\right\rangle}

\begin{titlepage}
\begin{center}

\hfill UT-16-17 \\
\vspace{2mm}
\hfill IPMU-16-0060


\vspace{3.0cm}
{\large\bf Models of 750 GeV quarkonium and the LHC excesses}

\vspace{1.0cm}
{\bf Koichi Hamaguchi}$^{(a,b)}$ and 
{\bf Seng Pei Liew}$^{(a)}$ 
\vspace{1.0cm}

{\it
$^{(a)}${Department of Physics, University of Tokyo, Tokyo 113-0033,Japan}\\
$^{(b)}$ Kavli Institute for the Physics and Mathematics of the Universe (Kavli IPMU), \\
University of Tokyo, Kashiwa 277--8583, Japan
}

\vspace{1cm}
\abstract{
We investigate models involving a vector-like quark ($X$), which forms a 750 GeV bound state and reproduces the observed diphoton signals at the LHC, in connection with other excesses in the LHC data. An exotic hypercharge of $-4/3$ is required to fit the signal cross section, which indicates that there is additional particle(s) that mediates the decay of $X$ in the full theory. We find that, introducing an SU(2) doublet vector-like quark of mass around 600 GeV in our UV-complete framework can accommodate not only the diphoton but also the on-Z excess (and potentially a slight excess in the monojet events). Our models also include a dark matter candidate. The most useful way to probe the models at the LHC is via monojet searches. The relic dark matter density is largely determined by coannihilation effects, and indirect detection of dark matter annihilation signals is the alternative and complementary probe of our models.
}
\end{center}
\end{titlepage}

\setcounter{footnote}{0}

\section{Introduction}
The ATLAS and CMS collaborations have recently analyzed the Large Hadron Collider (LHC) data at a center-of-mass energy of 13 TeV (run 2) and reported an excess of diphoton events at about 750 GeV~\cite{ATLAS-CONF-2015-081,CMS-PAS-EXO-15-004}.
The ATLAS shows a local (global) significance of 3.9$\sigma$ (2.0$\sigma$), while the CMS reported a local (global) significance of 2.8-2.9 $\sigma$ ($<1\sigma$) depending on the spin hypothesis recently~\cite{ATLAS-CONF-2016-018,CMS-PAS-EXO-16-018}.

Among many possible interpretations,
one of the most attractive scenarios is the 750 GeV quarkonium, i.e., a QCD bound state of heavy vector-like quark with a mass of about 375 GeV~\cite{1512.06670,1602.05539,1602.08100,1602.08819}. 
Unlike many other diphoton models, the necessary ingredients for the diphoton excess in this scenario is extremely simple; just the existence of the vector-like quark $X$ with a small width, $\Gamma_X\ll \Gamma(S\to \gamma\gamma)\simeq {\cal O}(\MeV)$, can lead to a diphoton resonance, and it does not require an additional 750 GeV singlet field nor new strong dynamics other than the Standard Model (SM) gauge group.
In particular, it was shown in Refs.~\cite{1602.08100,1602.08819} that the diphoton signal rate can be explained by a bound state of vector-like quarks $X$ which transforms as $({\bf 3}, {\bf 1}, -4/3)$ under the ${\rm SU}(3)\times {\rm SU}(2)\times {\rm U}(1)_Y$ of the SM gauge groups.


Although the setup above for the diphoton signal is very simple, it cannot be the full theory. In particular, the theory must allow the decay of $X$ to avoid the severe collider bound on the long-lived colored particle~\cite{1604.04520}. 
In Refs.~\cite{1602.08100,1602.08819}, the decay of $X$ into multi-jets (and missing energy) was discussed.
In particular, Ref.~\cite{1602.08100} studied the case that $X$ decays via  higher dimensional operators, assuming that the decay products contain a dark matter (DM) final state with a mass close to that of the $X$. It was shown that the current collider and cosmological bounds on $X$ can be evaded if the mass difference between $X$ and the DM is about $30\sim 50~\GeV$.

In this paper, we propose explicit renormalizable models to complete this scenario.
As the vector-like quark $X$ in this scenario has an exotic quantum number, it cannot have a renormalizable interaction with the SM quarks. 
Therefore, an additional colored particle must be introduced to mediate the decay of $X$. We investigate the possibility that such a colored mediator, which we denote by $Y$, is an SU(2) doublet vector-like quark with a SM quantum number $({\bf 3}, {\bf 2}, -5/6)$. The doublet consists of two quarks with charges $-4/3$ and $-1/3$, and it can couple to both the $X$ and the down-type SM quarks, mediating the decay of the $X$ field.

Interestingly, we have found that the decay of the lower component of the $Y$ into $X$ may also explain the ATLAS on-$Z$ excesses~\cite{1503.03290,ATLAS-CONF-2015-082}.\footnote{It is noted that the CMS collaboration has reported results consistent with SM predictions in the same channel. It could be due to the incompatibility between the collaborations at estimating the number of background events, as emphasized in Ref.~\cite{1601.05777}. In this work, the ATLAS excesses are interpreted as a contribution originated from BSM physics. See Refs.~\cite{1503.04184,1504.01768,1504.02752,1504.02244,1504.07869,1504.04390,1506.05799,1506.07161,1506.08803,1507.08471,1510.07691,1601.05777,1602.05075} for previous theoretical efforts interpreting the excess.} The on-$Z$ analyses are supersymmetry (SUSY) searches conducted to look for events with dilepton, jets and missing energy. At 8 (13) TeV, 29 (21) events are observed while the expected SM background events are 10.6 (10.3), corresponding to a deviation of 3.0 $\sigma$ (2.2 $\sigma$) from SM.
We show that the on-$Z$ excess can be explained assuming that the vector-like quarks have masses $m_X\simeq 375~\GeV$ and $m_Y\simeq 600-700~\GeV$ and the decay of $X$ contains a large missing energy.

In the second part of this paper, we propose two explicit models.
Since both of the constraints from the $X$ search and the on-$Z$ excess point to the existence of dark matter (or missing energy), we introduce a dark matter field in addition to the vector-like quarks $X$ and $Y$.
In the first model, we consider an inert doublet dark matter, while
in the second model, a singlet scalar dark matter is introduced together with yet another vector-like quark.
We investigate LHC constraints as well as the dark matter phenomenology of these models. It is worth noting that monojet searches are of particular significance in our mass-degenerate scenarios, coinciding with the observation that there is a slight (1$\sigma$) excess in the 13 TeV monojet signal region~\cite{ATLAS-CONF-2015-062}. 
Therefore, LHC anomalies (on-$Z$, monojet) other than the diphoton resonance may be incorporated within our UV-complete framework.

The rest of the paper is organized as follows. In Sec.~\ref{sec:diphoton}, we briefly review the scenario of the 750 GeV quarkonium, in particular the case with a vector-like quark $X$ with a hypercharge $Y_X=-4/3$.
In Sec.~\ref{sec:onZ}, we introduce a new SU(2) doublet vector-like quark $Y$ to complete the model, and investigate its implication on the ATLAS on-$Z$ excesses. In Sec.~\ref{sec:models}, we propose explicit models, and LHC constraints and dark matter phenomenology in each model are studied. Summary and discussion are given in Sec.~\ref{summary}. The details of the LHC analyses are described in Appendix \ref{app:analysis}, and some analytic formulas for models in Sec.~\ref{sec:models} are presented in Appendix \ref{appendix:formulas}.

\section{750 GeV quarkonium and the diphoton excess}
\label{sec:diphoton}

In this section, we briefly review the scenario of the 750 GeV quarkonium, which is a bound state of vector-like quarks $X$ with a mass of $m_X\simeq 375~\GeV$. 
When a pair of the $X$ quarks are produced near the threshold energy, they can form a color-singlet, $S$-wave bound state and the production cross section is enhanced. 
In Ref.~\cite{1602.08100}, the diphoton cross section is calculated for various hypercharges $Y_X$, assuming that $X$ is a color triplet and SU(2) singlet, and it was shown that the case of $|Y_X|=4/3$ explains the best the diphoton excess. In Ref.~\cite{1602.08819}, the cross section is calculated for various constituent particles; for spins $0$, $1/2$ and $1$, various color representations, and different electromagnetic charges. A color-triplet fermion with electric charge $-4/3$ is among the best candidates, together with other possibilities such as color-triplet scalar with charge $-4/3$ or $5/3$ and color-sextet scalar with a charge $-2/3$.

In this paper, we consider the case that the $X$ has a quantum number $({\bf 3}, {\bf 1}, -4/3)$ under the ${\rm SU}(3)\times {\rm SU}(2)\times {\rm U}(1)_Y$ of the SM gauge group.
In Refs.~\cite{1602.08100,1602.08819}, the $\gamma\gamma$ production cross section through the bound state is estimated as 
\begin{align}
\sigma(pp\to S_0(X\bar{X})\to \gamma\gamma) \simeq 
\begin{cases}
5.7~\text{fb~\cite{1602.08100}}\,,
\\
4.7~\text{fb~\cite{1602.08819}}\,,
\end{cases}
\end{align}
up to (large) uncertainties discussed there.\footnote{In Ref.~\cite{1602.08819}, the QCD binding potential is approximated to be Coulomb-like~\cite{1204.1119}. Ref.~\cite{1602.08100} makes use of a different QCD potential, and includes the effects of the electric force.}

The  decay width of the 750 GeV bound state in this scenario is around 6 MeV~\cite{1602.08100}, indicating a narrow-width scalar resonance.
This setup predicts other bound state signals (dijet, diboson etc) as well, but they are thus far not strongly constrained. It should be noted that the the decay rate of the constituent fermion $X$ should be smaller than that of the bound state in order to enhance the diphoton signal.  This amounts to the condition $\Gamma_X \lesssim {\cal O}(1) \MeV$, which, as will be discussed in the following sections, can be easily satisfied in our models.

\section{Vector-like quarks and ATLAS on-$Z$ excess}
\label{sec:onZ}

As mentioned in the introduction, the setup of the previous section cannot be the complete model and $X$ should have an additional interaction which allows its decay. 
In Refs.~\cite{1602.08100,1602.08819}, the decay of $X$ into multi-jets (and missing energy) was discussed
without specifying the UV completion of the model.
In this paper, we propose explicit renormalizable models to complete the scenario.
As in Ref.~\cite{1602.08100}, we assume that the decay of $X$ contains a DM in the final state.

Because of the exotic charge $Y_X=-4/3$, the $X$ quark cannot have a renormalizable interaction with the SM field. Therefore, a new colored field should be introduced to mediate its decay. We consider a vector-like fermion for such a new colored mediator and denote it by $Y$. Assuming that $X$ and $Y$ are coupled via a Yukawa coupling with the SM Higgs, $Y$ should be an SU(2) doublet with a hypercharge $-5/6$ or $-11/6$. 
We consider the former case, since it can couple to the SM down-type quarks while the latter case needs further additional colored fields, i.e., 
\begin{align}
Y=
\begin{pmatrix}
B'\\ X'
\end{pmatrix}_{-5/6},
\end{align}
where $B'$ and $X'$ has electric charges $-1/3$ and $-4/3$, respectively.
The interaction is then given by
\begin{align}
-{\cal L}_{\text{int}} &= (\bar{Y} H) (\lambda_X + \lambda_{X5}\gamma_5) X
+ h.c.
\label{eq:L_XYH}
\end{align}
where $H=(H^+,H^0)^T$ is the SM Higgs doublet and $(\bar{Y} H)=\bar{B'} H^+ + \bar{X'} H^0$.
Here and hereafter, we assume $\lambda_{X5}=0$ for simplicity.
As we shall see in the explicit models discussed in the next section, 
the $Y$ field can couple to the SM quarks and the DM field directly or indirectly, allowing the decay of $X$ field through the $Y$.  

Before discussing the details of the models in the next section, let us discuss the possible explanation of the ATLAS on-$Z$ excess in this scenario. 
The interaction in \eqref{eq:L_XYH} causes a mixing between $X$ and $X'$, resulting in mass eigenstates $X_1$ and $X_2$. The heavier one can decay into the lighter one emitting a $Z$ boson or the Higgs boson:
\begin{align}
X_2\to X_1 + Z/h\,.
\end{align}
As we will see, this decay channel leads to a signal with $Z+$ jets $+$ missing energy, which may explain the ATLAS on-$Z$ excess.

The major features of the ATLAS on-$Z$ analysis are as follows. The final state must include at least a pair of opposite sign same flavor (OSSF) leptons (electron or muon) with invariant mass close to the $Z$ boson mass. The required missing transverse energy is $\MET > 225~\GeV$. At least two signal jets (reconstructed using the anti-$k_t$ algorithm~\cite{0802.1189}) are required, and a cut of $H_{\text{T}} > 600~\GeV$, where $H_{T}$ is the scalar sum of jet $p_T$, is applied to the final state (see Appendix~\ref{app:analysis} for further details).
Assuming that the number of observed events in the signal region follows the Poisson distribution, and by adding the statistical uncertainty as well as the systematic uncertainty in quadrature, the number of excessive events due to new physics are:
\bea
N_{\rm BSM} &= 18.4 \pm 6.3\quad \text{in the Run 1 analysis},
\\
N_{\rm BSM} &= 10.7 \pm 5.1\quad \text{in the Run 2 analysis}.
\eea

We have used the following pipeline to estimate the on-$Z$ excess fit of our model:
{\tt  MadGraph 5 v2.2.3}~\cite{Alwall:2014hca,Alwall:2011uj} for event generation, {\tt Pythia 6.4}~\cite{Sjostrand:2006za} for parton shower and {\tt Checkmate}~\cite{1312.2591,1503.01123} for analysis implementations. {\tt Checkmate} utilizes {\tt Delphes 3}~\cite{deFavereau:2013fsa} (which has {\tt FastJet} incorporated~\cite{Cacciari:2011ma,Cacciari:2005hq}) for detector simulations. 

We have generated three samples ($pp \to X_2\overline{X}_2  \to X_1\overline{X}_1+ZZ/Zh/hh$) and made appropriate rescaling to vary the branching ratio of $X_2  \to X_1 Z$ while fitting the on-$Z$ excess. In addition, the process $pp \to B'\overline{B'}  \to X_1\overline{X}_1+W^+W^-$ is also generated and included in our analysis. Our model is implemented using {\tt Feynrules v2.3.1}~\cite{1310.1921}, and the vector-like quark pair production cross section is estimated  up to NNLO accuracy using {\tt Hathor v2.0}~\cite{1007.1327}. 

\begin{figure}[t]
\centering
\includegraphics[scale=0.6]{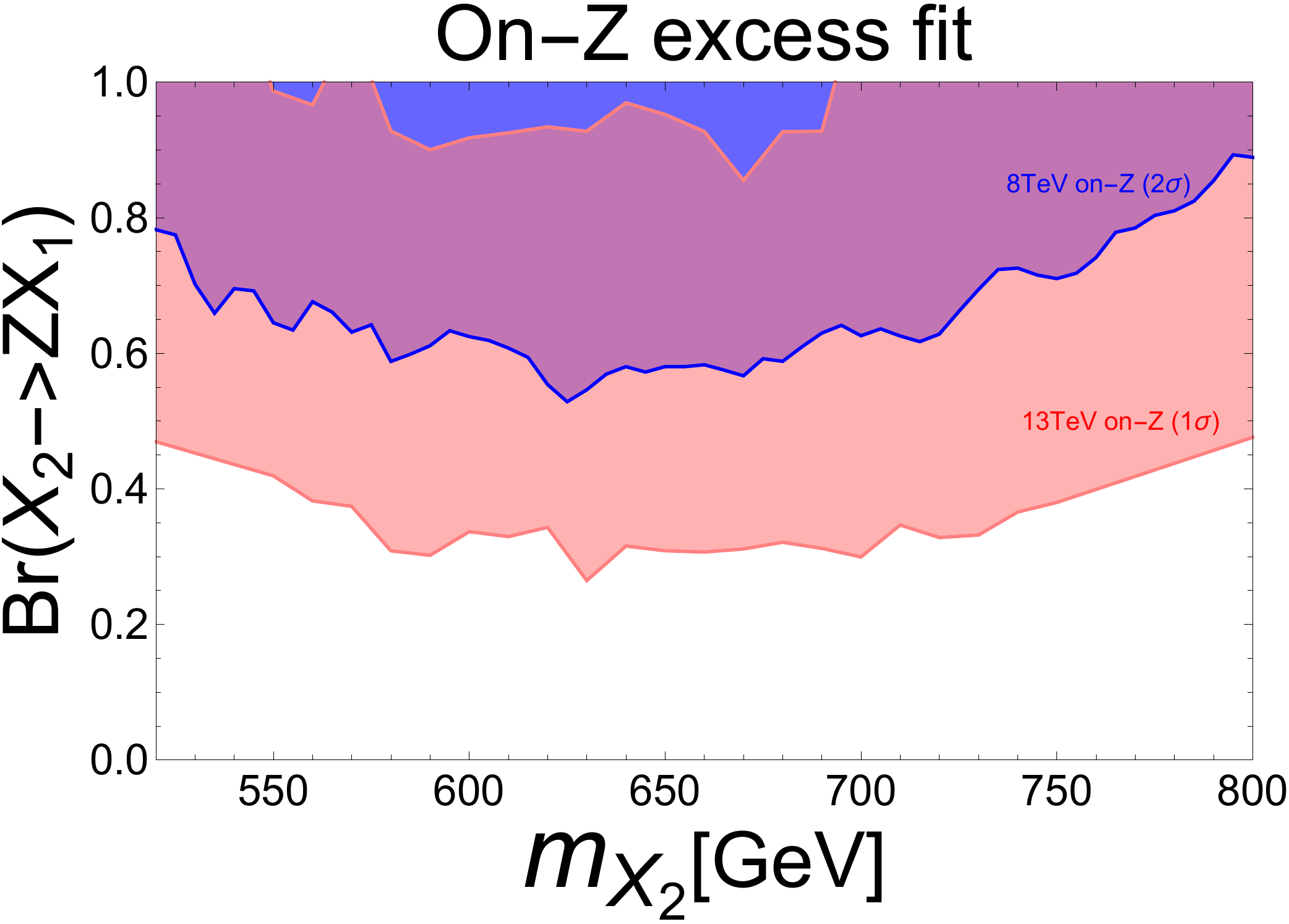}
\caption{The on-$Z$ excess fit on the $\text{Br}(X_2  \to X_1 Z)$ and $X_2$ mass plane
for $m_{X_1}=375~\GeV$. The blue region shows the fit of Run 1 data at statistical significance of 2 $\sigma$, while the pink band indicates the fit of Run 2 data at 1 $\sigma$.}
\label{fig:fit}
\end{figure}

The result of our fit is shown in Fig.~\ref{fig:fit}. Here, $X_1$ is assumed to decay into DM ($\MET$) accompanied by two jets, while the mass splitting between $X_1$ and DM is set to be $40~\GeV$. Jets from the decay of $X_1$ are relatively soft and the on-$Z$ result is insensitive to the details of $X_1$ decay.
It can be seen that $\text{Br}(X_2  \to X_1 Z) \gtrsim 0.6$ is necessary in order to fit the Run 1 on-$Z$ excess at 2 $\sigma$. The requirement to fit the Run 2 excess is looser, with $\text{Br}(X_2  \to X_1 Z) \gtrsim 0.4$ at 1 $\sigma$. The reason is twofold, one: the production cross section is enhanced at 13 TeV; two: the observed number of excessive events is smaller in Run 2.

\begin{figure}
\centering
\begin{subfigure}{.5\textwidth}
  \centering
  \includegraphics[width=1.1\linewidth]{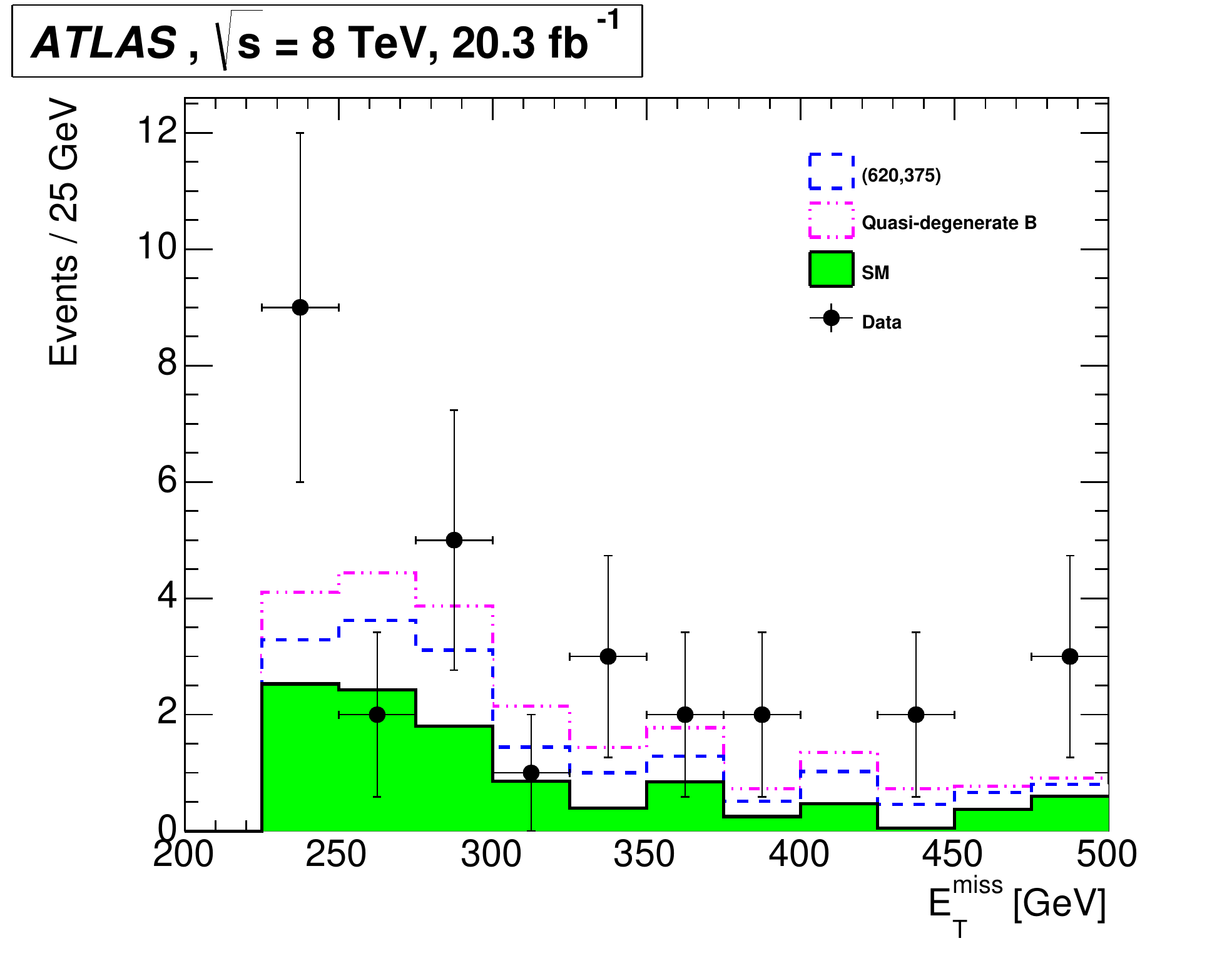}
\end{subfigure}%
\begin{subfigure}{.5\textwidth}
  \centering
  \includegraphics[width=1.1\linewidth]{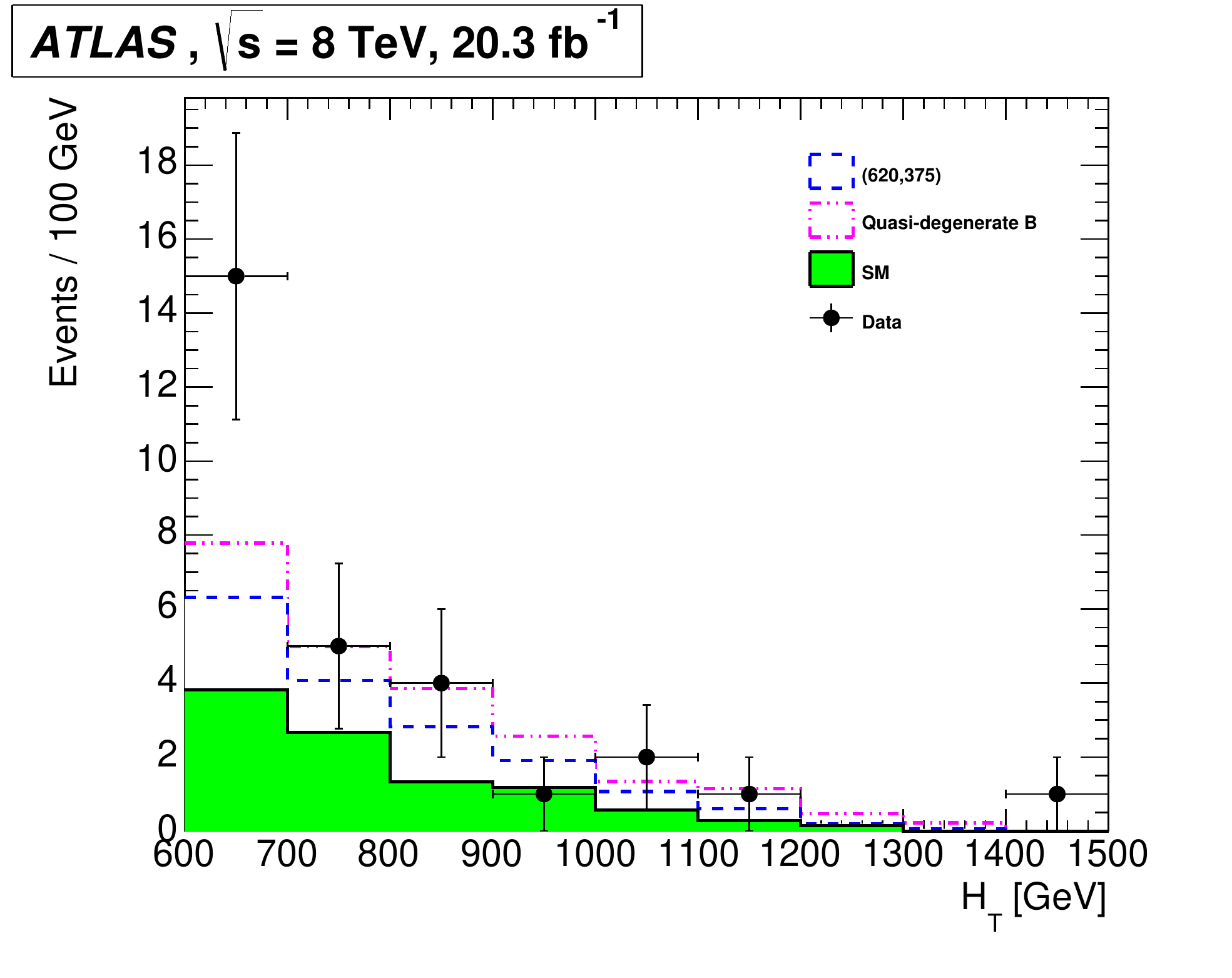}
\end{subfigure}
\begin{subfigure}{.5\textwidth}
  \centering
  \includegraphics[width=1.1\linewidth]{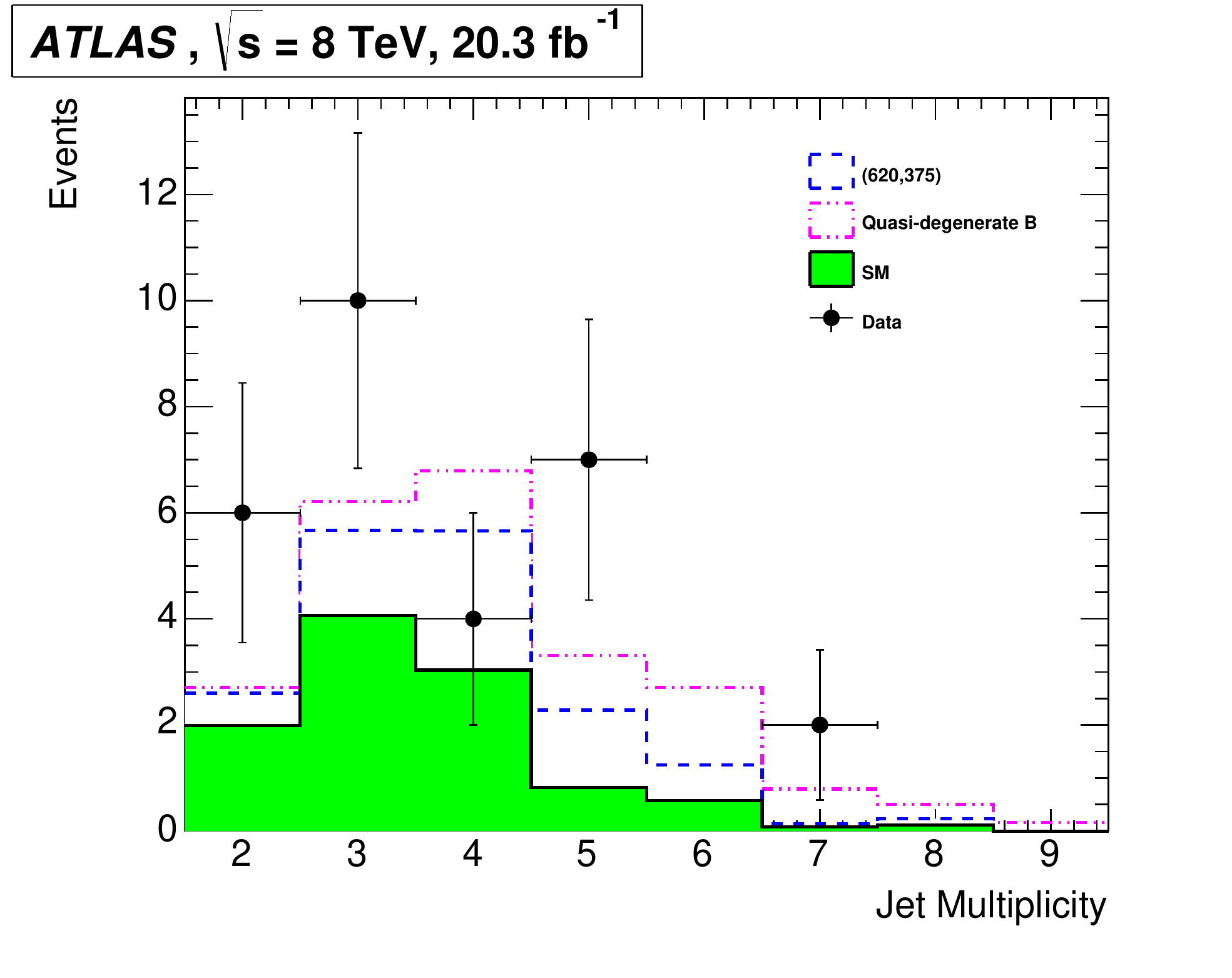}
\end{subfigure}
\caption{\textit{(from top left to bottom)} The distribution of $\MET$, $H_{\text{T}}$ and jet multiplicity for Run 1 (8 TeV), for the model point with 
$m_{X_2}=620~\GeV$, $m_{X_1}=375~\GeV$, and $\text{Br}(X_2\to X_1 + Z)=0.8$ (blue)
and the ``quasi-degenerate $B$" scenario (pink), on top of the expected SM background (green).
}
\label{fig:dis8}
\end{figure}
\begin{figure}
\centering
\begin{subfigure}{.5\textwidth}
  \centering
  \includegraphics[width=1.1\linewidth]{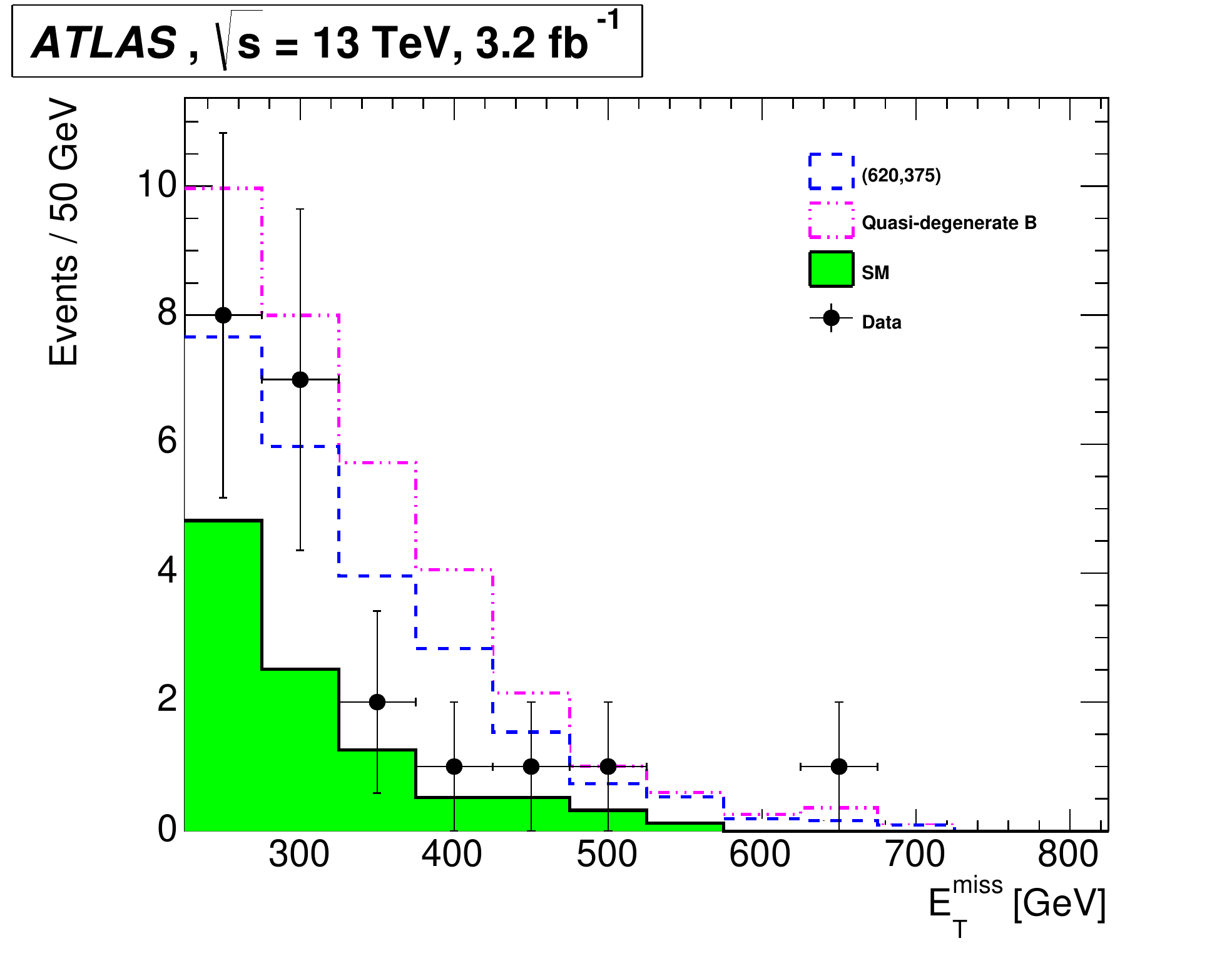}
\end{subfigure}%
\begin{subfigure}{.5\textwidth}
  \centering
  \includegraphics[width=1.1\linewidth]{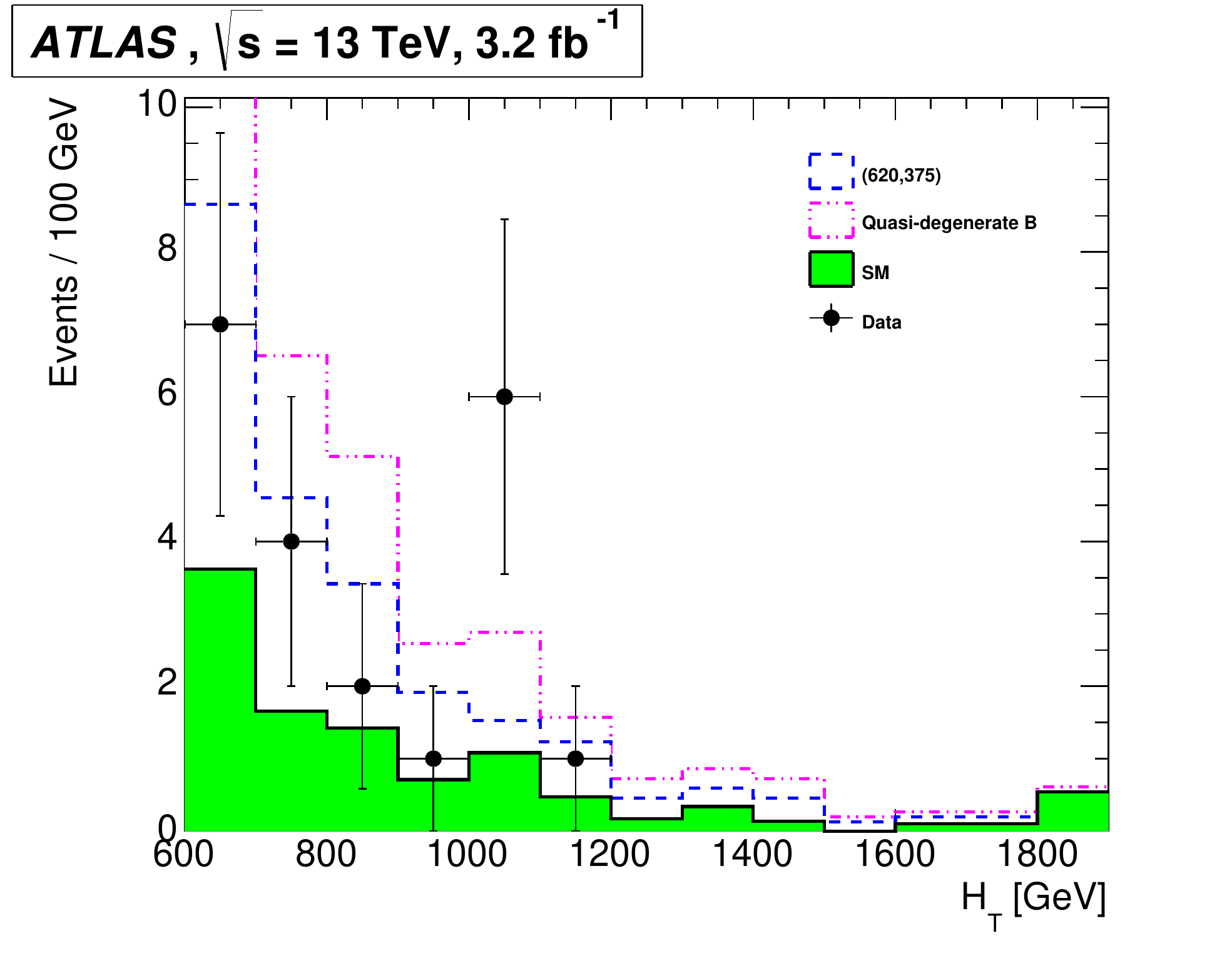}
\end{subfigure}
\begin{subfigure}{.5\textwidth}
  \centering
  \includegraphics[width=1.1\linewidth]{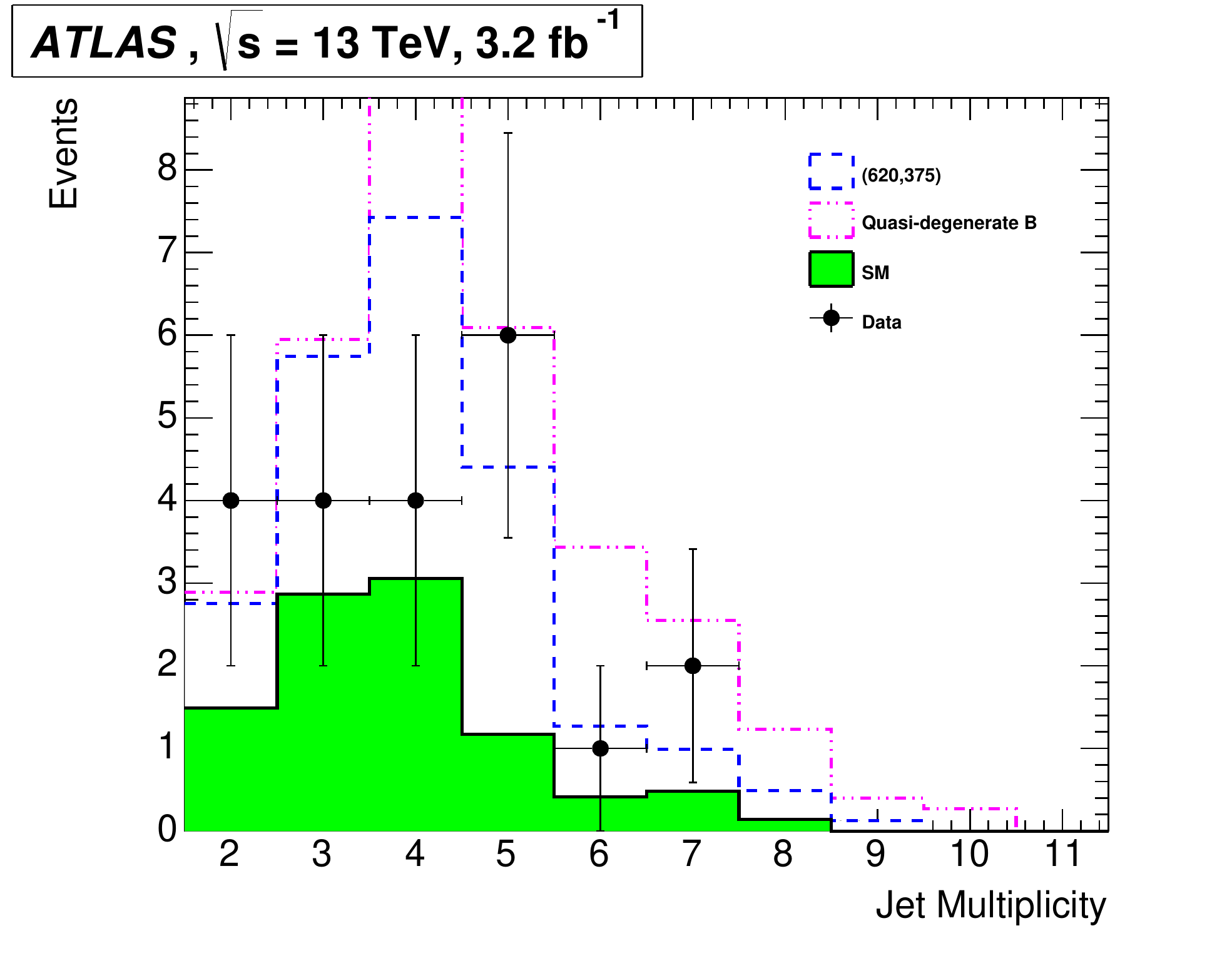}
\end{subfigure}
\caption{\textit{(from top left to bottom)} The distribution of $\MET$, $H_{\text{T}}$ and jet multiplicity for Run 2 (13 TeV), for the model point with 
$m_{X_2}=620~\GeV$, $m_{X_1}=375~\GeV$, and $\text{Br}(X_2\to X_1 + Z)=0.8$ (blue)
and the ``quasi-degenerate $B$" scenario (pink), on top of the expected SM background (green).
}
\label{fig:dis13}
\end{figure}

In Figures~\ref{fig:dis8} and~\ref{fig:dis13}, we show the predicted $\MET$, $H_{\text{T}}$ and jet multiplicity distribution for a benchmark point 
on top of expected SM background as compared to the Run 1 and Run 2 data, respectively.
As the parameters of the benchmark point, we have chosen $m_{X_2}=620~\GeV$, $m_{X_1}=375~\GeV$ and $\text{Br}(X_2\to X_1 + Z)=0.8$.\footnote{This point corresponds to $(m_X,m_Y)\simeq (440,560)~\GeV$ and $\lambda_X\simeq 0.6$. See Section~\ref{sec:models} and Figure~\ref{fig:BrXYeta} for details.}
The blue dashed line shows the contributions from $X_2\to X_1+Z/h$ (and $B'\to X_1+W$),
which produces $N_{\rm BSM}= 8.5~(13.4)$ events at 8 (13) TeV. 
As shown in Fig.~\ref{fig:fit}, the contribution at 8 TeV is fitted up to 2 $\sigma$ compared with the excess, while the fit is reasonably better at 13 TeV (within 1 $\sigma$).

As an illustration, we also show the distributions for another benchmark point,``quasi-degenerate $B$", by pink dashed lines. Here, we have an additional particle $B$ or $B_2$ in the mass eigenstate, with $m_{B_2}=705~\GeV$ (see Section~\ref{sec:heavyb} for details of model building).  $B_2$ decays solely to $WX_2$, as other decay channels are kinematically closed.
This ``quasi-degenerate $B$" scenario produces $N_{\rm BSM}= 10.7~(17.2)$ events at 8 (13) TeV, which gives more signal events than the benchmark point discussed above for both Run 1 and Run 2 data.

We have also checked other relevant collider analyses which potentially constrain our model. It is found that the multi-jet plus $\MET$ search~\cite{1405.7875} gives the most stringent constraint, though our benchmark scenarios are still allowed. Furthermore, the direct pair production of $X_1$ can be constrained by monojet searches. This constraint will be further discussed in the next Section.

\section{Models for 750 GeV diphoton and on-$Z$ excesses}
\label{sec:models}

In this section, we introduce two explicit models which can explain the 750 GeV diphoton and on-$Z$ excesses simultaneously. 
The matter contents of the models are summarized in Table.~\ref{table:matter_contents}.

\begin{table}[t]
\begin{tabular}{|l|l||l|l|}
\hline
model $XY\eta$ & $(SU_3,SU_2)_{U_{1Y}}$ & 
model $XYB\phi$ & $(SU_3,SU_2)_{U_{1Y}}$ 
\\ \hline
$X$ & $({\bf 3},{\bf 1})_{-4/3}$ & 
$X$ & $({\bf 3},{\bf 1})_{-4/3}$
\\
$Y=(B',X')^T$ & $({\bf 3},{\bf 2})_{-5/6}$ &
$Y=(B',X')^T$ & $({\bf 3},{\bf 2})_{-5/6}$
\\
$\eta=(\eta^+,\eta^0)^T$ & $({\bf 1},{\bf 2})_{1/2}$ &
$B$ & $({\bf 3},{\bf 1})_{-1/3}$
\\
&&
$\phi$ & $({\bf 1},{\bf 1})_0$
\\ \hline
\end{tabular}
\caption{Matter contents of the two models. $X$, $Y$, and $B$ are Dirac
fermions, $\eta$ is a complex scalar, and $\phi$ is a real scalar. All fields are odd under $Z_2$ parity.}
\label{table:matter_contents}
\end{table}

\begin{itemize}
\item
In the first model, we introduce an inert doublet $\eta$ in addition to $X$ and $Y$.
We assume $m_Y>m_X>m_\eta$.
\item
In the second model, a singlet scalar dark matter $\phi$ is introduced together with yet another vector-like quark $B$. We consider two cases:
\begin{itemize}
\item[(i)] $m_Y > m_X >  m_B > m_\phi$ (light $B$),
\item[(ii)] $m_B > m_Y  > m_X > m_\phi$ (heavy $B$).
\end{itemize}
\end{itemize}

Generically, our models contain a colored mediator which couples to DM as well as SM quarks. Such quark generation-dependent couplings cannot be arbitrary due to stringent constraints on the first and second quark generation couplings from flavor changing neutral current (FCNC) processes~\footnote{An example of the FCNC calculations with similar setup can be found in Ref.~\cite{1403.0324}.}. We choose the couplings to the third generation to be the dominant one in order to avoid these constraints.

\subsection{Model $XY\eta$}
\label{sec:inert}

From the matter content in Table~\ref{table:matter_contents},
 the most generic renormalizable Lagrangian is
\begin{align}
{\cal L} &= {\cal L}_{\text{SM}}
+ \sum_{F=Y,X}\bar{F}(i\slashed{D}-m_F)F
+|D_{\mu}\eta|^2-m_{\eta}^2|\eta|^2
 -V(H,\eta)
\nonumber\\
&
- (\bar{Y} H) (\lambda_X + \lambda_{X5}\gamma_5) X
-y_{\eta i}\overline{d_{Ri}}P_L(Y\cdot \eta)
+ h.c.
\label{eq:Lagrangian_Intert}
\end{align}
where 
$(\bar{Y} H)=\bar{B'} H^+ + \bar{X'} H^0$,
$(Y\cdot \eta)=B' \eta^0 - X' \eta^+$,
and $d_{Ri}$ denotes the SM right-handed down-type quarks.
For simplicity, we assume $\lambda_{X5}=0$.
We further assume that $Y$ mainly couples to the third generation quark, and take $y_{\eta 1}=y_{\eta 2}=0$ and $y_{\eta 3}=y_\eta$.
The Yukawa couplings $\lambda_X$ and $y_\eta$ 
can be taken to be real and positive by field redefinitions.
As mentioned in the previous section, 
the $X$ and $X'$ are mixed through the Yukawa interaction in \eqref{eq:Lagrangian_Intert},
leading to mass eigenstates $X_1$ and $X_2$.
The main decay modes of the new particles are given by, assuming $m_{X_2}\simeq m_{B'}>m_{X_1}>m_\eta$
and $\lambda_X\gg y_\eta$, 
\begin{align}
X_2 &\to \begin{cases} X_1 Z \\ X_1 h \end{cases},
\quad
B' \to X_1 W,
\quad
X_1\to b \eta^-,
\quad
\eta^- \to \eta^0 + \pi^-\,.
\end{align}
The mixings and decay rates of the mass eigenstate quarks are summarized in Appendix.~\ref{appendix_model_inert}.

For simplicity, 
we consider the case where there is no renormalizable interaction between $\eta$ and the SM Higgs. The loop corrections make the neutral component slightly lighter than the charged component, i.e., $\Delta m\simeq 350\MeV$~\cite{hep-ph/0512090}. Then, the decay of $\eta^-$ produces only very soft pion with a decay rate~\cite{hep-ph/9804359}
\begin{align}
\Gamma(\eta^-\to \eta^0+ \pi^-) \simeq 3 \times 10^{-14} \GeV \lrfp{\Delta m}{350\MeV}{3}\beta\,,
\end{align}
where $\beta=\sqrt{1-(m_{\pi^-}/\Delta m)^2}$. Therefore, the $\eta^-$ can be regarded as a missing particle at collider searches.

\subsubsection{Diphoton and ATLAS on-$Z$ excesses}
\begin{figure}[t]
\centering
\includegraphics[scale=1.0]{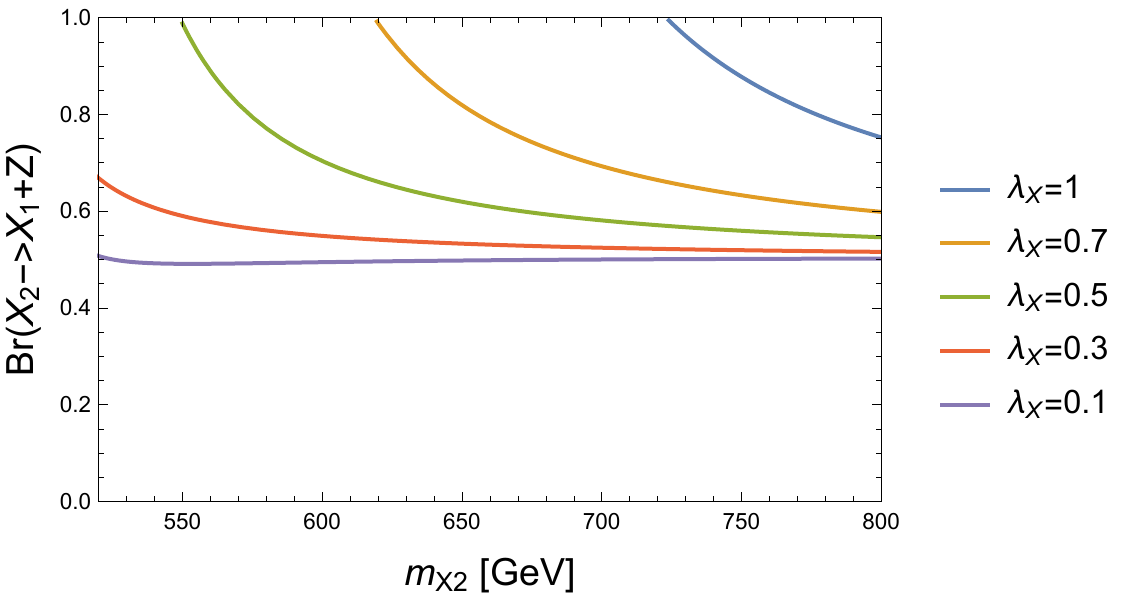}
\caption{The branching fraction $\text{Br}(X_2\to X_1 Z)$ 
as a function of $m_{X_2}$, for $m_{X_1}= 375~\GeV$, $\lambda_X=0.1-1$, and $y_\eta\ll \lambda_X$.}
\label{fig:BrXYeta}
\end{figure}

As discussed in Sec.~\ref{sec:diphoton} and \ref{sec:onZ}, this model may explain the diphoton and ATLAS on-$Z$ excesses simultaneously. 
\begin{itemize}

\item
The decay rate of the $X_1$ is given by (cf. Appendix.~\ref{appendix_model_inert})
\begin{align}
\Gamma(X_1\to b\eta^-) &\simeq 1.1\times 10^{-7}~\GeV\lrfp{\lambda_X}{0.1}{2}\lrfp{y_\eta}{0.01}{2}
\end{align}
where we have taken $(m_{X_2},m_{X_1},m_\eta)\simeq (620, 375, 325)~\GeV$ as a benchmark point. 
Therefore, it can easily satisfy the condition for the diphoton excess, $\Gamma(X_1\to b\eta^-)\ll \Gamma(S_0(X\bar{X})\to \gamma\gamma)\simeq 1~\MeV$, for $y_\eta\ll \lambda_X\lesssim 1$.

\item
For the on-$Z$ excess, in Fig.~\ref{fig:BrXYeta} we show the branching fraction $\text{Br}(X_2\to X_1 Z)$
as a function of $m_{X_2}$, for $m_{X_1}= 375~\GeV$, $\lambda_X=0.1-1$, and $y_\eta\ll \lambda_X$.
Compared with Fig.~\ref{fig:fit}, one can see that both the Run 1 and Run 2 excesses can be explained in a large region of parameter space.\footnote{The region close to $\text{Br}(X_2\to X_1 Z)=1$ corresponds to the maximal mixing, $\sin 2\theta_X\simeq 1$, where $m_X\simeq m_Y\simeq (m_{X_1}+m_{X_2})/2$. 
(See Appendix.~\ref{appendix_model_inert}.)}

\end{itemize}

\subsubsection{Other LHC constraints}

\begin{figure}[t]
\centering
\includegraphics[scale=0.7]{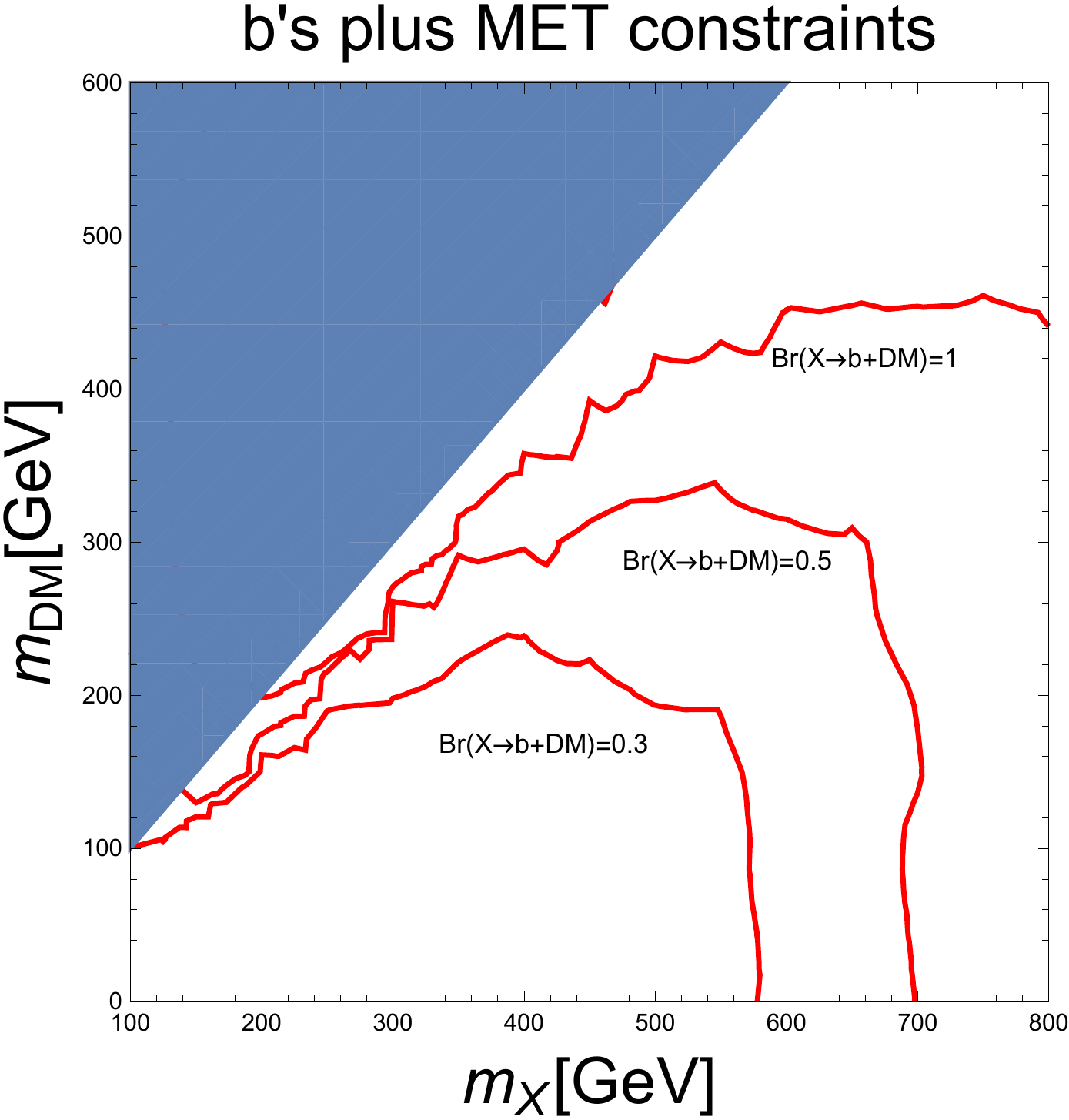}
\caption{$b$-jets plus $\MET$ constraints recast from ATLAS sbottom search~\cite{1308.2631} for different choices of decay branching ratio. The signal cross section has been normalized to that of the vector-like fermion pair production cross section calculated using {\tt Hathor}.}
\label{fig:bmet}
\end{figure}

Since the 375 GeV vector-like quark $X_1$ decays to $\eta^- b$ with $\eta^-$ decaying into a very soft pion, this decay mode could lead to final state with $b$-jets plus large $\MET$. We make use of the acceptance times efficiency table provided by the ATLAS sbottom search~\cite{1308.2631} to estimate its constraints on this channel. The result is shown in Fig.~\ref{fig:bmet}. As can be seen in Figure~\ref{fig:bmet}, models with large $X_1-\eta$ mass splitting ($\gtrsim 60~\GeV$) are disfavored.
 
The situation where $X_1$ decays to a soft jet and an almost mass degenerate $\eta$ could avoid this collider bound, but may be constrained by monojet searches~\cite{1407.0608,1502.01518,ATLAS-CONF-2015-062}, which are relevant when $X_1$ is produced in pair and recoiled against an energetic initial state radiation (ISR). In order to study the monojet constraints, again we generate samples of monojet events with {\tt MadGraph} plus {\tt Pythia}. Jet-parton matching is performed using the MLM scheme~\cite{hep-ph/0206293,hep-ph/0611129,0706.2569}. We use the parameter $r_{\rm obs}\equiv (N_{\rm sig}-1.96\Delta N_{\rm sig})/N^{95\%}_{\rm obs}$ to estimate the 95\% exclusion limits on our model (the model is considered excluded if $r_{\rm obs}>1$), with $N_{\rm sig}$ being the number of BSM events in the signal region, $\Delta N_{\rm sig}$ the uncertainty, $N^{95\%}_{\rm obs}$ the observed 95 \% limit on the event number given by the experimental collaborations.

\begin{figure}[t]
\centering
\includegraphics[scale=0.7]{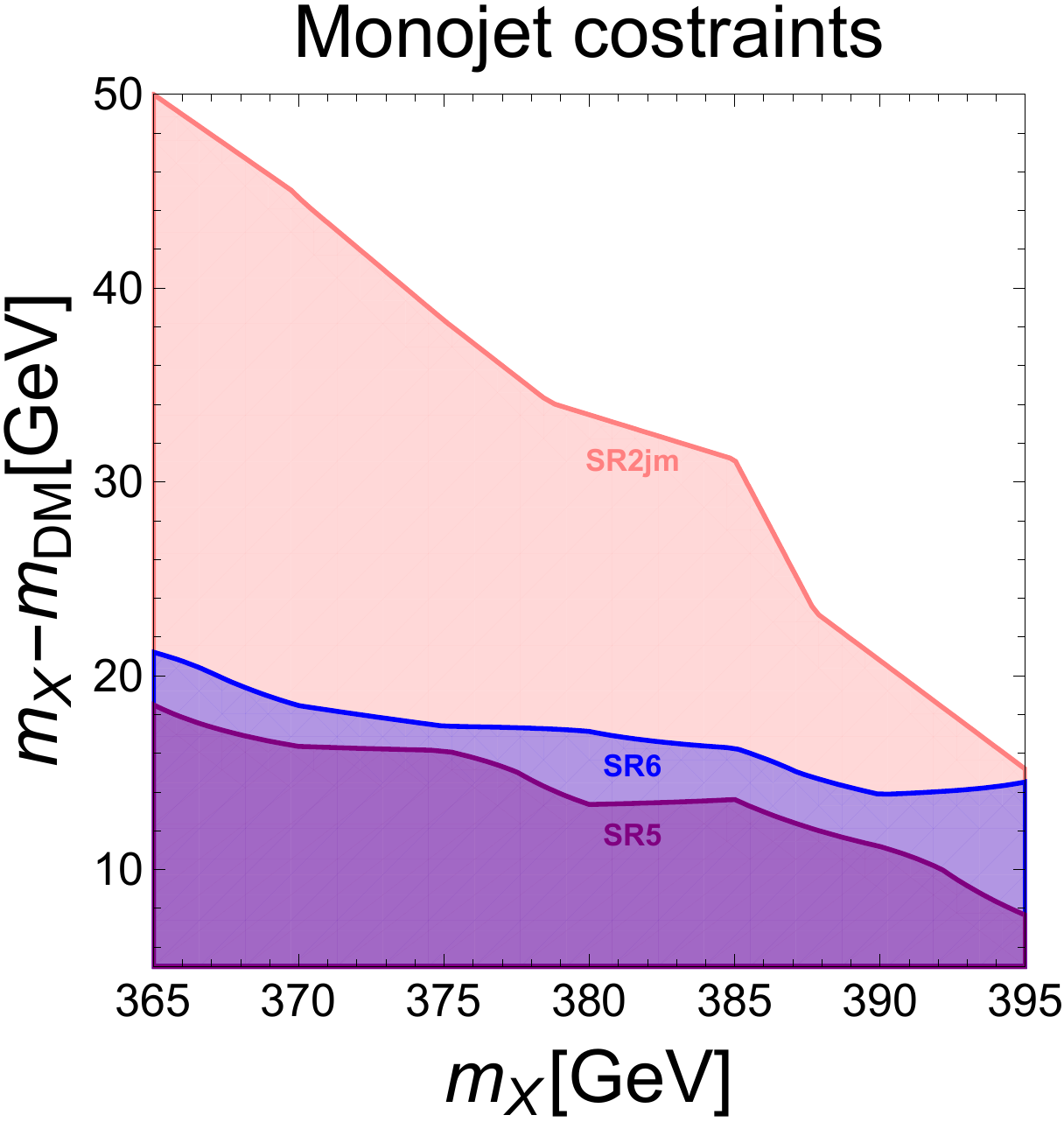}
\caption{Monojet constraints on the $XY\eta$ model. The most senstive signal regions are SR2jm (pink)~\cite{ATLAS-CONF-2015-062}, SR6 (blue) and SR5 (purple)~\cite{1502.01518}. Below the thick line of each signal region, the parameter region is excluded.}
\label{fig:mono_inert}
\end{figure}

Our results are presented on the $X_1-\eta$ mass splitting versus $X_1$ mass plane as in Figure~\ref{fig:mono_inert}. We find that the ATLAS 13 TeV monojet search (signal region SR2jm) gives the most stringent constraint~\cite{ATLAS-CONF-2015-062}, followed by signal regions SR6 and SR5 of the 8 TeV monojet analysis~\cite{1502.01518}. Our model (with $m_{X_1}\simeq 375~\GeV$) is still viable for mass splitting larger than 40 GeV. Finally, we note that currently there is a slight excess of around $1\sigma$ in SR2jm. It is exciting to point out that our model could as well accommodate the excess, though the excess is appreciably less significant than that of diphoton and on-$Z$. This channel should definitely be studied rigorously in the future runs.

\subsubsection{Dark Matter}

The properties of $\eta$ are completely determined by SM gauge symmetry as we consider the case where there is no renormalizable interaction between $\eta$ and the SM Higgs. Since there is no direct couplings to quarks or gluons, it is difficult to probe this model with direct detection experiments.

On the other hand, the s-wave annihilation cross section is given as~\cite{hep-ph/0512090}:
\bea
\sigma v \simeq 9 \times 10^{-26} \text{cm}^3\text{s}^{-1} \left(\frac{m_{\rm DM}}{335~\GeV}\right)^{-2},
\eea
while the search on dwarf spheroidal galaxies (dSphs) for gamma rays with six years of Fermi-LAT data places a 95\% confidence level limit of $\sigma v \lesssim 8.5\times 10^{-25} \text{cm}^3\text{s}^{-1}$ at 335 GeV after taking dSphs DM density profile uncertainties into account~\cite{1503.02641}. We expect improvements of this bound in the near future with more data taking. Note that we need non-thermal mechanism to achieve the observed relic DM density as the DM annihilation cross section is too large.

\subsection{Model $XYB\phi$ with light $B$}
\label{sec:lightb}
In this subsection, we consider the $XYB\phi$ model in Table~\ref{table:matter_contents},
assuming $m_Y > m_X >  m_B > m_\phi$.
From the matter content in the table,
 the most generic renormalizable Lagrangian is
\begin{align}
{\cal L} &= {\cal L}_{\text{SM}}
+ \sum_{F=Y,X,B}\bar{F}(i\slashed{D}-m_F)F
+ \frac{1}{2}(\partial \phi)^2  - \frac{1}{2}m_\phi^2 \phi^2
-V(H,\phi)
\nonumber\\
&
- (\bar{Y} H) (\lambda_X + \lambda_{X5}\gamma_5) X
- \overline{B}(\lambda_B + \lambda_{B5}\gamma_5) (Y\cdot H)
- y_{\phi i}\; \phi \overline{B}P_R d_{Ri}
+ h.c.
\label{eq:Lagrangian_B}
\end{align}
where 
$(\bar{Y} H)=\bar{B'} H^+ + \bar{X'} H^0$ and
$(Y\cdot H)=B' H^0 - X' H^+$.
For simplicity, we assume $\lambda_{X5}=\lambda_{B5}=0$.
We further assume $y_{\phi 1}=y_{\phi 2}=0$ and denote $y_{\phi 3}=y_\phi$.
The Yukawa couplings $\lambda_X$, $\lambda_B$ and $y_\phi$ 
can be taken real and positive by field redefinitions.
We also assume that the scalar interaction $V(H,\phi)$ does not affect DM abundance.
The Yukawa couplings in \eqref{eq:Lagrangian_B} causes mixings between $B$ and $B'$, as well as between $X$ and $X'$, leading to mass eigenstates $B_{1,2}$ and $X_{1,2}$. 
The main decay channels are given by, assuming $m_{X_2}\simeq m_{B_2}>m_{X_1}>m_{B_1}>m_\phi$
and $y_\phi \ll 1$,
\begin{align}
X_2 &\to \begin{cases} B_1 W \\ X_1 Z \\ X_1 h   \end{cases},
\quad
B_2 \to \begin{cases} X_1 W \\ B_1 Z \\ B_1 h   \end{cases},
\quad
X_1\to B_1 W^*,
\quad
B_1\to b \phi\,.
\end{align}
The mixings and decay rates of the fields are summarized in Appendix \ref{appendix_model_B}.

\subsubsection{Diphoton and ATLAS on-$Z$ excesses}
As discussed in Sec.~\ref{sec:diphoton} and \ref{sec:onZ}, this model may explain the diphoton and ATLAS on-$Z$ excesses simultaneously. 
\begin{itemize}
\item
The decay rate of the $X_1$ is, using {\tt Madgraph}~\cite{1402.1178}, estimated as, for a benchmark point $(m_Y,m_X,m_B)\simeq (620,375,360)~\GeV$,
\begin{align}
\Gamma(X_1\to B_1 +W^*\to B_1+\ell\nu/jj)\sim 10^{-6}~\GeV \cdot \theta_X^2 \theta_B^2,
\end{align}
with $\theta_X$ and $\theta_B$ being the mixing angles of the vector-like quarks (see Appendix~\ref{appendix:formulas}). Therefore, it satisfies the condition for the diphoton excess, $\Gamma(X_1\to b\eta^-)\ll \Gamma(S_0(X\bar{X})\to \gamma\gamma)\simeq 1~\MeV$, while avoiding the collider constraints for long-lived particle. 

\item For the on-$Z$ excess, let us first consider the case $\lambda_B\ll \lambda_X$. In this case $X_2$ and $B_2$ mainly decays into $X_1+Z/h$ and $X_1 + W$, respectively. The branching fraction $\text{Br}(X_2\to X_1 Z)$ is essentially the same as the $XY\eta$ model discussed in Sec.~\ref{sec:inert}. (See Fig.~\ref{fig:BrXYeta}.) Thus, the on-$Z$ excesses in Run 1 and Run 2 can be explained in a large region of parameter space. For $\lambda_B\sim \lambda_X$, the branching fraction $\text{Br}(X_2\to X_1 + Z/h)$ decreases, but $\text{Br}(B_2\to B_1 + Z/h)$ increases instead. The sum of them is close to unity up to kinematical corrections, and hence the result does not depend on much on the hierarchy between $\lambda_X$ and $\lambda_B$.

\end{itemize}

\subsubsection{Other LHC constraints}

We have also studied monojet bounds on the model $XYB\phi$ with light $B$. The monojet bounds are significantly stronger than the $XY\eta$ model
since we now have a light $B$ produced in pair recoiled against ISR as well. As a result, we do not find any parameter region that could avoid such constraints. To show this, we fix $m_{X_1}=375~\GeV$, and choose two benchmarks: $(m_{\rm DM},m_B)=(320,365), (305, 365)$, where the numbers are in unit GeV. The first benchmark corresponds to a parameter point where the thermal relic DM density is achieved via coannihilation (see Section~\ref{sec:dm}). The second benchmark corresponds to a point where the monojet constraint is found to be the least stringent. Note that the $m_B-m_{\rm DM}$ mass  splitting of this benchmark is $60~\GeV$. The monojet constraints are further relaxed beyond $60~\GeV$ but such a parameter region is excluded by the $b$-jets plus $\MET$ search (cf. Figure~\ref{fig:bmet}). The results are shown in Table~\ref{table2}.  Again, it should be noted that there is a slight excess in the 13 TeV monojet signal region. Should the excess grow or persist in future LHC runs, this model would become relevant.   
\begin{table}[t]
\begin{tabular}{|c|c|c|c|c|}
\hline
$(m_{\rm DM},m_{B})$ [GeV]& SR & Bkg & Obs & $R_{\rm obs}$
\\ \hline
\multirow{3}{*}{(320,365)}
&2jm
&$163\pm 20$
&186
&1.77
\\
&SR5
&$8300\pm 300$
&7988
&0.79
\\
&M2
&$8620\pm 270$
&8606
&0.62
\\ \hline
\multirow{3}{*}{(305,365)}
&2jm
&$163\pm 20$
&186
&1.45
\\
&SR5
&$8300\pm 300$
&7988
&0.66
\\
&M2
&$8620\pm 270$
&8606
&0.48
\\ \hline
\end{tabular}
\caption{Benchmarks for the $XYB\phi$ model with light $B$. The number of background (Bkg) and observed (Obs) events, as well as $R_{\rm obs}$ of the most sensitive signal region of each of the monojet analyses are shown~\cite{1407.0608,1502.01518,ATLAS-CONF-2015-062}. The first benchmark corresponds to a parameter point where the thermal DM relic density matches the observed value ($\Omega_{\rm DM} h^2\simeq 0.12$). The second benchmark is a parameter point where, within our exploration, the monojet constraints are found to be least stringent.}
\label{table2}
\end{table}

\subsubsection{Dark Matter}
\label{sec:dm}

\begin{figure}[]
\centering
\includegraphics[scale=0.6]{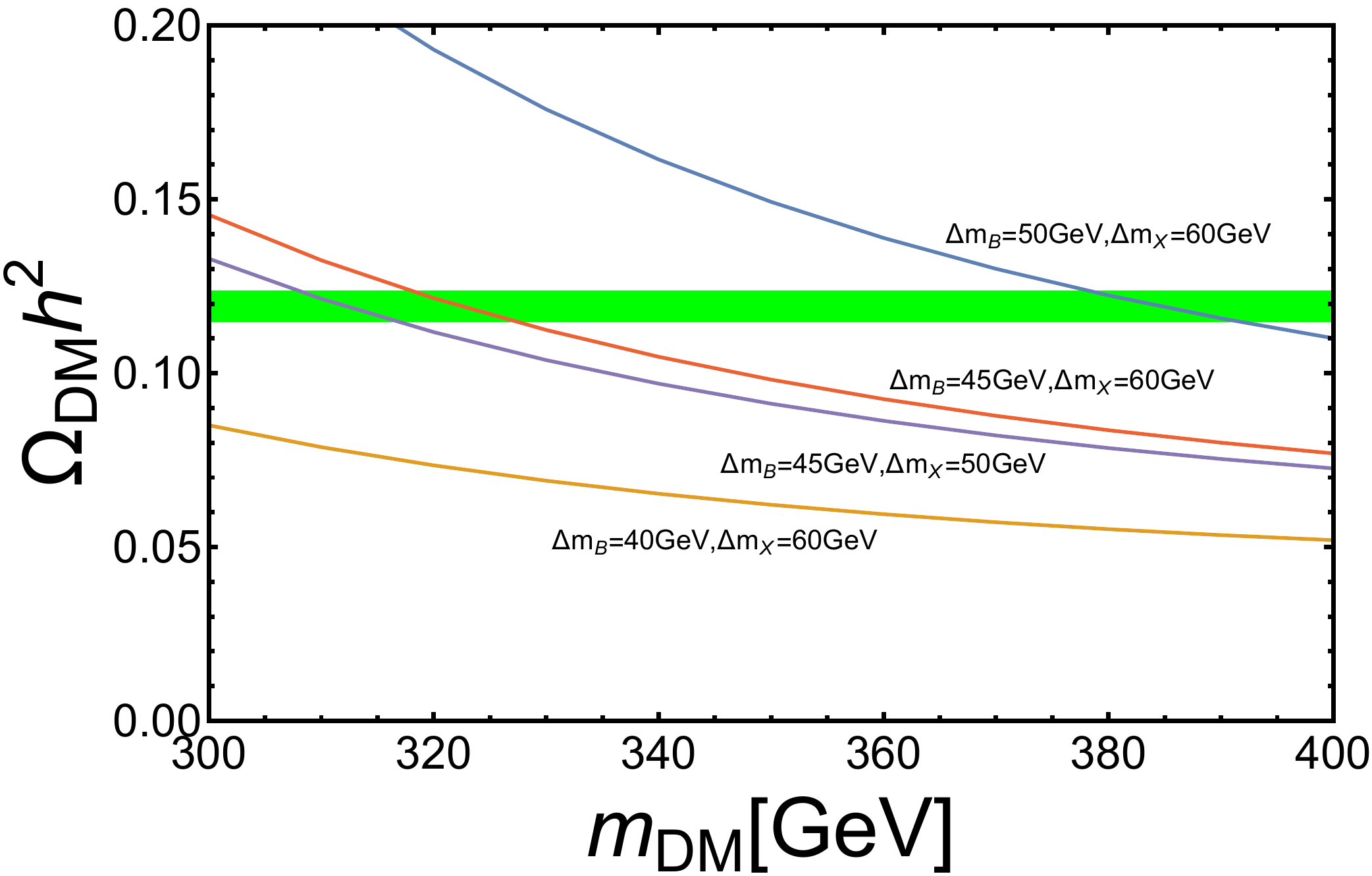}
\caption{$\Omega_{\rm DM}h^2$ versus DM mass. $\Delta m_{X}$ ($\Delta m_{B}$) is the mass difference between DM and $X$ ($B$).}
\label{fig:omega}
\end{figure}

DM-DM annihilation occurs via t-channel mediators and suffers from helicity suppression. The relic density of DM is instead determined by the annihilation rate of $X_1$ and $B$. This effect is characterized by the effective annihilation cross section~\cite{Griest:1990kh}:
\bea
\sigma_{\rm eff} v=\sum_{i=X_1,B}\sigma_i v \left(\frac{n_i^{eq}}{n^{eq}}\right)^2,
\eea
where $n_i^{eq}$ is the number density of particle $i$ in thermal equilibrium, while $n^{eq}$ is the sum of the number density of $X_1, B$ and DM in thermal equilibrium. $\sigma_i v$ is the self annihilation cross section of the fermion pairs, which is dominated by the QCD annihilation into a pair of gluons or quarks. We integrate the Boltzmann equation numerically to obtain the thermal relic DM density. In Figure~\ref{fig:omega}, we show the thermal relic DM density for some choices of mass splitting parameters. The parameter set $(m_{\rm DM},m_B, m_{X_1})\simeq(320,365,375)\GeV$ reproduces the observed relic DM density in the universe.

Let us briefly comment on the direct detection prospects of the model. As DM couples only to the bottom quark via $B$, the scattering of DM off a nucleon is only via loop-level interaction between DM and the nucleon's gluons. We found that the constraint on $y_\phi$ from the LUX experiment~\cite{1512.03506}  is of $O(10)$ for $B$ mass range of several hundred GeV to a few TeV. Probing the model with direct detection is extremely difficult.

\subsection{Model $XYB\phi$ with quasi-degenerate or heavy $B$}
\label{sec:heavyb}
Finally, we consider the scenario where the Lagrangian is the same as \eqref{eq:Lagrangian_B}, but with mass relations $m_B\gtrsim m_Y > m_X  > m_\phi$. As we shall see, the minimal model would lead to a long lived $X_1$ at the collider scale, and we need additional new particles to make $X_1$ decay promptly.

The model where $B$ is quasi-degenerate with $Y$ produces more on-$Z$ signal events
than the case where $B$ is much heavier than $Y$ or the inert scalar doublet scenario in Section~\ref{sec:inert}, as explained in Section~\ref{sec:onZ}. Meanwhile,  the $X_1$-DM mass splitting has to be small to avoid jets plus large $\MET$ constraints. As a result, two-body or three-body decay modes are kinematically closed within this minimal framework. $X_1$ can only decay via the following four-body decay mode: $X_1\to W^*B_1^*\to jjj\phi$. However, the decay rate of this process is $\Gamma_{\rm 4-body} \lesssim 10^{-12}\GeV \cdot \theta_X^2 \theta_B^2$, which means that the colored and electrically charged $X_1$ typically decays outside the detector, leading to stringent collider constraints~\cite{1604.04520}. One is then forced to introduce new particle(s) to make $X_1$ decay promptly. One relatively simple possibility is to, instead of $\phi$, consider fermion DM $\chi$ along with a color triplet scalar $\tilde{u}$ of quantum number $({\bf 3}, {\bf 1}, 2/3)$. Assume the following interaction:
\bea
\Delta{\cal  L}=\epsilon_{abc} y_X X^a \tilde{u}^b u^c + \delta_{ab} y_{\chi}\tilde{u}^{*a}u^b \chi + h.c.,
\eea
where we have written down the color indices of the fields explicitly. The three-body decay rate is, based on dimensional analysis, $\Gamma_{X}\sim (y_X^2y_{\chi}^2/128\pi^3)(m_{X}^5/m_{\tilde{u}}^4)\sim O(1) \MeV$ for $O(1)$ Yukawa couplings and $m_{\tilde{u}}=1\TeV$. The formation of 750 GeV bound state requires $\Gamma_{X} \lesssim 1 \MeV$, which translates to $m_{\tilde{u}}\gtrsim1\TeV$ for $O(1)$ Yukawa couplings.\footnote{Note that, in this setup, the $Y$ field does not play the role of mediating the $X$ decay. 
The decay of $X$ occurs through new scalar $\tilde{u}^{*a}$, while
the ATLAS on-$Z$ excess is explained by the decays of $Y$ as well as $B$.}

Let us briefly discuss other aspects of the $\chi$ model. Monojet constraints have been discussed in Ref.~\cite{1602.08100}, where the $X_1$-DM mass splitting has to be $\gtrsim 30~\GeV$. Let us assume that $\tilde{u}$ couples mainly to the top quark. The self annihilation cross section of $\chi$ is, for $y_{\chi}\sim O(1)$ and $m_{\tilde{u}}\gtrsim 1 \TeV$,  $\sigma \simeq (y_{\chi}^4/32\pi)(m_t^2/m_{\tilde{u}}^4)\lesssim 3 \times 10^{-27}  \text{cm}^3\text{s}^{-1}$, which is one order of magnitude smaller than the canonical thermal annihilation cross section. However, the self annihilation of the slightly heavier $X_1$ can help reduce the relic density of $\chi$. A mass splitting of around 40 GeV would lead to the correct value of the observed abundance of dark matter in the universe.   The prospect of direct detection is slim as $\chi$ couples to the third-generation quark. Indirect detection may be feasibe using the Cherenkov Telescope Array (CTA)~\cite{Wood:2013taa}. Finally, the TeV-scale colored $\tilde{u}$ would be produced copiously at the LHC, and can be probed using standard techniques of SUSY searches.

\section{Summary and Discussion}
\label{summary}

Motivated by the LHC diphoton excess, we have considered some models involving a 750 GeV quarkonium, 
which is a bound state of vector-like quarks $X$, in connection with dark matter and other excesses observed at the LHC. 
An exotic hypercharge of $Y_X=-4/3$ is required to fit the diphoton signal, indicating that there is additional particle(s) that mediates the decay of $X$ in the full theory. We have introduced an SU(2) doublet vector-like quark of mass around 600 GeV, and showed that not only the diphoton but also the on-Z excess (and potentially a slight excess in the monojet events) can be accommodated in our models.

Let us summarize the major findings of this paper. The three scenarios proposed here have distinctive properties, but all of them can accommodate the diphoton and on-$Z$ excess. The scenario with inert scalar doublet (Section~\ref{sec:inert}) has inert scalar mass around 40 GeV below the 375 GeV diphoton-to-be vector-like fermion $X$. As the annihilation cross section of DM is relatively large, non-thermal mechanism is needed to reproduce the correct DM relic density. On the flip side, the DM annihilation signals can optimistically be probed in near future. The second scenario (Section~\ref{sec:lightb}) proposed in the paper involves the cascade decay of $X$ into an almost degenerate bottom-like fermion, $B$, which in turn decays to DM. However, this model predicts too many monojet events. Even so, as there is currently a slight excess in the 13 TeV monojet channel, this model should be given attention if the monojet excess persists or grows in the next run. The correct relic DM density is obtained via coannihilations with $X$ and $B$. The third scenario (Section~\ref{sec:heavyb}) has $B$ heavier than $X$. At certain mass region, the model can provide more on-$Z$ signal events than the previous scenarios. However, in the minimal setup, $X$ is unable to decay promptly. One needs to introduce a colored scalar triplet ($\tilde{u}$) to mediate the decay efficiently. This model may be tested with CTA by looking for the DM annihilation signals, or probed directly by searching for $\tilde{u}$ collider signatures at the LHC. The observed relic DM density is achieved via $X$ coannihilation.

The observed diphoton excess has a relatively simple interpretation, i.e. a QCD bound state of heavy colored particles. We have argued that the full theory of such an interpretation can have rich collider and dark matter phenomenology. While the full theory necessitates the introduction of additional particles, we are encouraged by the fact these particles can provide ingredients to accommodate other LHC anomalies (on-$Z$, monojet). New experimental data, particularly from the next run of the LHC and dark matter indirect detection experiments (e.g. Fermi-LAT) will definitely test our proposal very soon.

\section*{Acknowledgments}
The authors thank Feng Luo, Michihisa Takeuchi, and Yoshitaro Takaesu for useful discussions. SPL is supported by JSPS Research Fellowships for Young Scientists and the Program for Leading Graduate Schools, MEXT, Japan. 
This work was supported by Grant-in-Aid for Scientific research 
(Nos.\ 26104001, 26104009, 26247038, 26800123, 16H02189)
and by World Premier International
Research Center Initiative (WPI	 Initiative), MEXT, Japan.

\appendix

\section{Experimental Analyses}
\label{app:analysis}

In this appendix, we describe the details of LHC analyses important to our study. 

We begin with the ATLAS SUSY searches for final states containing an on-shell $Z$ (on-$Z$)\cite{1503.03290,ATLAS-CONF-2015-082}. Both 8 TeV and 13 TeV ATLAS on-$Z$ analyses have the following requirements/cuts:
\begin{itemize}
  \item requires at least a pair of OSSF leptons with $81\GeV<m_{ll}<101~\GeV$
  \item $\MET > 225~\GeV$
  \item Number of jets $n_j \geq2$
  \item $H_{\text{T}} > 600~\GeV$
  \item azimuthal distance between jets and $\slashed{p}_T$ $\Delta\phi(j,\slashed{p}_T)>0.4$
\end{itemize}

There are slight differences in the requirement of the $p_T$ ordered $i$-th leptons ($p^{l_i}_T$) and jets ($p^{j_i}_T$). For the 8 TeV analysis:
\begin{itemize}
  \item $p^{l_1}_T>25~\GeV$, $p^{l_2}_T>10-14\GeV$ depending on trigger
  \item $p^{j}_T>35~\GeV$
\end{itemize}
For the 13 TeV analysis:
\begin{itemize}
  \item $p^{l_1}_T>50~\GeV$, $p^{l_2}_T>25~\GeV$
  \item $p^{j}_T>30~\GeV$
\end{itemize}

Monojet searches have been performed by the ATLAS collaboration to study SUSY with compressed mass spectra~\cite{1407.0608} and dark matter production at the LHC~\cite{1502.01518} at Run 1. For~\cite{1407.0608}, the following cuts are applied to all signal regions:
\begin{itemize}
\item requires at least a jet with $p_T>150~\GeV$, i.e. $p^{j_1}_T>150~\GeV$
\item $n_j \leq 3$ for jets with $p_T>30~\GeV$
  \item $\MET > 150~\GeV$
  \item $\Delta\phi(j,\MET)>0.4$ for all jets
  \item no leptons with $p_T>10~\GeV$
\end{itemize}
Events passing these cuts are further divided into three signal regions M1-M3 which have additional requirements on $\MET$ and $p_T^j$. For the dark matter production monojet search~\cite{1502.01518}:
\begin{itemize}
\item $p^{j_1}_T>150~\GeV$
\item $p^{j_1}_T/\MET>0.5$ 
  \item $\MET > 150~\GeV$
  \item $\Delta\phi(j,\MET)>1.0$ for jets with $p_T>30~\GeV$
  \item no leptons with $p_T>7\GeV$
\end{itemize}
There are 9 signal regions (SR1-9) corresponding to further different cuts on $\MET$  applying on events passing cuts above.

Furthermore, the Run 2 SUSY search for squarks and gluino is placing strong monojet limits~\cite{ATLAS-CONF-2015-062}. The relevant signal region is SR2jm and has kinematic cuts as follows:
\begin{itemize}
\item $p^{j_1}_T>300~\GeV$
\item $p^{j_2}_T>50~\GeV$
  \item $\MET > 200~\GeV$
  \item $\Delta\phi(j,\MET)>0.4$ up to three leading jets in the events
  \item no leptons with $p_T>10~\GeV$
  \item $\MET/\sqrt{H_T} > 15~\GeV^{1/2}$
    \item $m_{\rm eff} > 1600~\GeV$
\end{itemize}
 where $m_{\rm eff}$ is defined as the sum of $H_T$ and $\MET$. We summarize signal regions sensitive to our models in Table~\ref{tab:sr}. 
 
 \begin{table}[t]
	\centering
	{\small
	\begin{tabular}{l|ccccccc|r}
	& $\MET$\,[GeV] & $p_{T\,j_1}$\,[GeV] & $\Delta\phi\,(j, \slashed{p}_T)$ & $n_j$
	& $H_T$\,[GeV] & $\MET/\sqrt{H_T}$ & $m_{\rm eff}$\,(incl.)  & Ref. \\
	\hline
	on-$Z$ (8TeV) & 225 & 35 & 0.4 & $\ge 2$& 600 & - & -  & \cite{1503.03290} \\
	on-$Z$ (13TeV) & 225 & 30 & 0.4 & $\ge 2$& 600 & - & -  & \cite{ATLAS-CONF-2015-082} \\
	\hline
	M2 & 340 & 340 & 0.4 & $\le 3$ & - & - & -& \cite{1407.0608} \\
	SR5 & 350 & 0.5$\MET$ & 1.0 & - & - & - & - & \cite{1502.01518} \\
	SR6 & 400 & 0.5$\MET$ & 1.0 & - & - & - & - & \cite{1502.01518} \\
	SR2jm & 200 & 300 & 0.4 & $\ge 2$  & - & 15\,GeV$^{1/2}$ & 1.2\,TeV & \cite{ATLAS-CONF-2015-062} \\
	\hline
	\end{tabular}
	}
	\caption{Signal regions most sensitive to our studies. Notations are defined in text.}
	\label{tab:sr}
\end{table}

\section{Mixings and decay rates of vector-like quarks}
\label{appendix:formulas}
In this appendix, we describe the mixings and decay rates of vector-like quarks 
in the models discussed in Sec.~\ref{sec:models}.

\subsection{Model $XY\eta$}
\label{appendix_model_inert}

Assuming $\lambda_{X5}=0$, the mass terms of $(X',X)$ in \eqref{eq:Lagrangian_Intert} become
\begin{align}
-{\cal L}_{\text{mass}}
&=
(\overline{X'},\overline{X})
\begin{pmatrix}
m_Y & \lambda_X v
\\
\lambda_X v & m_X
\end{pmatrix}
\begin{pmatrix}
X'\\X
\end{pmatrix}
\end{align}
where $v=\vev{H^0}\simeq 174~\GeV$. It can be diagonalized by
\begin{align}
\begin{pmatrix}
X'\\X
\end{pmatrix}
=
\begin{pmatrix}
\cos\theta_X & -\sin\theta_X
\\
\sin\theta_X & \cos\theta_X
\end{pmatrix}
\begin{pmatrix}
X_2\\X_1
\end{pmatrix}
\end{align}
where
\begin{align}
\theta_X = \frac{1}{2}\arctan\lrf{2\lambda_X v}{m_Y-m_X}
\simeq 
\frac{\lambda_X v}{m_Y-m_X}
\simeq
0.07\lrf{\lambda_X}{0.1}\lrf{250~\GeV}{m_Y-m_X}
\end{align}
The mass splitting in the doublet, $m_{X_2}-m_{B'}$, is small for $\lambda_X\ll 1$.
\begin{align}
m_{X_2}-m_{B'} 
&\simeq \frac{(\lambda_X v)^2}{m_Y-m_X}
\simeq 1.2~\GeV
\lrfp{\lambda_X}{0.1}{2}
\lrf{250~\GeV}{m_Y-m_X}.
\end{align}

The partial decay rates of $X_2$ are
\begin{align}
\Gamma(X_2\to X_1 Z)&=
\frac{1}{64\pi}
\cos^2\theta_X \sin^2\theta_X
\frac{g_2^2 m_{X_2}^3}{m_W^2}
f_V\left(r_1, r_Z\right)
\overline{\beta}\left(r_1, r_Z\right)
\\
\Gamma(X_2\to X_1 h)&=
\frac{1}{32\pi}
(\cos 2\theta_X)^2
\lambda_X^2 m_{X_2}
f_h\left(r_1, r_h\right)
\overline{\beta}\left(r_1, r_h\right)
\\
\Gamma(X_2\to \eta^- b)&=
\frac{1}{32\pi}\cos^2\theta_X y_\eta^2 m_{X_2}
\left(1-r_\eta^2+r_b^2\right)
\overline{\beta}\left(r_\eta,r_b\right)
\end{align}
where  $r_1=m_{X_1}/m_{X_2}$, $r_Z=m_Z/m_{X_2}$, $r_h=m_h/m_{X_2}$, 
$r_\eta=m_\eta/m_{X_2}$, and $r_b=m_b/m_{X_2}$.
The functions $f_V$, $f_h$, and $\overline{\beta}$ are given by
\begin{align}
f_V(r_1,r_V) &=(1-r_1^2)^2 + (1-6 r_1 + r_1^2) r_V^2 - 2r_V^4\,,
\label{eq:fV}
\\
f_h(r_1,r_h) &= 1 + 2r_1 + r_1^2 - r_h^2\,,
\\
\overline{\beta}(a,b) &=
\sqrt{1 +a^4 +b^4 -2a^2 - 2b^2 - 2a^2b^2}\,.
\label{eq:beta}
\end{align}
In the limit of $m_Y-m_X\gg v$ and $\lambda_X\gg y_\eta$, 
the branching fractions become
$\text{Br}(X_2\to X_1 Z)\simeq \text{Br}(X_2\to X_1 h)\simeq 0.5$,
as expected from the Goldstone equivalence theorem.

The decay rate of the $X_1$ is given by
\begin{align}
\Gamma(X_1\to \eta^- b) &= 
\frac{1}{32\pi}\sin^2\theta_X y_\eta^2 m_{X_1}
\left(1-r_{\eta 1}^2+r_{b 1}^2\right)
\overline{\beta}\left(r_{\eta 1},r_{b 1}\right)
\end{align}
where $r_{\eta 1}=m_\eta/m_{X_1}$ and $r_{b 1}=m_b/m_{X_1}$.
Finally, the decay rate of the $\eta^-$ is given by~\cite{hep-ph/9804359}
\begin{align}
\Gamma(\eta^-\to \eta^0 \pi^-) &= \frac{G_F^2}{\pi}{\rm cos}^2\theta_c f_{\pi}^2 \Delta m^3 \sqrt{1-m_{\pi^-}^2/\Delta m^2}.
\end{align}
$G_F$ is the Fermi constant. $\text{cos }\theta_c$ is the Cabibo angle. $f_{\pi}$ and $m_{\pi^-}$ are the decay constant and mass of pion respectively. $\Delta m$ is the mass splitting between $\eta^-$ and $\eta^0$.

\subsection{Model $XYB\phi$}
\label{appendix_model_B}

Assuming $\lambda_{X5}=\lambda_{B5}=0$, the mass terms of $(X',X)$ and $(B',B)$ in \eqref{eq:Lagrangian_B} become
\begin{align}
-{\cal L}_{\text{mass}}
&=
\sum_{F=X,B}
(\overline{F'},\overline{F})
\begin{pmatrix}
m_Y & \lambda_F v
\\
\lambda_F v & m_F
\end{pmatrix}
\begin{pmatrix}
F'\\F
\end{pmatrix}
\end{align}
where $v=\vev{H^0}\simeq 174~\GeV$. They can be diagonalized by
\begin{align}
\begin{pmatrix}
F'\\F
\end{pmatrix}
=
\begin{pmatrix}
\cos\theta_F & -\sin\theta_F
\\
\sin\theta_F & \cos\theta_F
\end{pmatrix}
\begin{pmatrix}
F_2\\F_1
\end{pmatrix}
\quad (F=X,B)
\end{align}
where
\begin{align}
\theta_F = \frac{1}{2}\arctan\lrf{2\lambda_F v}{m_Y-m_F}
\simeq 
\frac{\lambda_F v}{m_Y-m_F}
\simeq
0.07\lrf{\lambda_F}{0.1}\lrf{250~\GeV}{m_Y-m_F}
\end{align}
The mass splitting in the doublet, $m_{X_2}-m_{B_2}$, is small for $\lambda_{B,F}\ll 1$.
\begin{align}
m_{F_2}-m_Y 
&\simeq \frac{(\lambda_F v)^2}{m_Y-m_F}
\simeq 1.2~\GeV
\lrfp{\lambda_F}{0.1}{2}
\lrf{250~\GeV}{m_Y-m_F}
\quad (F=X,B).
\end{align}

The partial decay rates of $X_2$ are
\begin{align}
\Gamma(X_2\to B_1 W)&=
\frac{1}{32\pi}
\cos^2\theta_X \sin^2\theta_B
\frac{g_2^2 m_{X_2}^3}{m_W^2}
f_V\left(r_{B_1}, r_W\right)
\overline{\beta}\left(r_{B_1}, r_W\right)
\\
\Gamma(X_2\to X_1 Z)&=
\frac{1}{64\pi}
\cos^2\theta_X \sin^2\theta_X
\frac{g_2^2 m_{X_2}^3}{m_W^2}
f_V\left(r_{X_1}, r_Z\right)
\overline{\beta}\left(r_{X_1}, r_Z\right)
\\
\Gamma(X_2\to X_1 h)&=
\frac{1}{32\pi}
(\cos 2\theta_X)^2
\lambda_X^2 m_{X_2}
f_h\left(r_{X_1}, r_h\right)
\overline{\beta}\left(r_{X_1}, r_h\right)
\end{align}
where  $r_{B_1}=m_{B_1}/m_{X_2}$, $r_W=m_W/m_{X_2}$,
$r_{X_1}=m_{X_1}/m_{X_2}$,  $r_Z=m_Z/m_{X_2}$,  and $r_h=m_h/m_{X_2}$.
The functions $f_V$, $f_h$, and $\overline{\beta}$ are given in Eqs.~\eqref{eq:fV}-\eqref{eq:beta}.
The partial decay rates of $B_2$ are obtained by replacing $X$ and $B$.
In the limit of $m_Y-m_{X,B}\gg v$, the branching fractions become
$\text{Br}(X_2\to B_1 W)\simeq C_B/(C_X+C_B)$ and
$\text{Br}(X_2\to X_1 Z)\simeq \text{Br}(X_2\to X_1 h)\simeq (1/2)\cdot C_X/(C_X+C_B)$
where $C_F=\lambda_F^2(1+r_F)^2(1-r_F^2)$ ($F=B,X$),
as expected from the Goldstone equivalence theorem.

\bibliographystyle{aps}
\bibliography{ref}

\end{document}